\documentclass[%
 reprint,
 amsmath,amssymb,
 aps,
]{revtex4-2}

\usepackage{graphicx}
\usepackage{dcolumn}
\usepackage{bm}
\usepackage{array} 
\usepackage{braket}
\begin{document}

\preprint{APS/123-QED}

\title{Does Cosmology require Hermiticity in Quantum Mechanics?}

\author{Oem Trivedi}
\email{oem.trivedi@vanderbilt.edu}
\author{Alfredo Gurrola}
\email{alfredo.gurrola@vanderbilt.edu}
\affiliation{Department of Physics and Astronomy, Vanderbilt University, Nashville, TN, 37235, USA}

\date{\today}

\begin{abstract}
We explore the consequences of allowing non-Hermitian structures in quantum cosmology by extending the Wheeler–DeWitt framework beyond strictly Hermitian dynamics. Using a controlled semiclassical reduction, we show how anti-Hermitian contributions propagate into both early universe primordial fluctuations and late-time structure growth as effective damping or gain terms. Confronting this framework with inflationary observables, growth of structure and the observed near flatness of the universe, we derive strong infrared constraints that suppress non-Hermiticity across cosmic history. We demonstrate that these bounds are mutually consistent between early and late-time probes and can be partially relaxed in theories beyond General Relativity. Our results establish cosmology as a novel arena for testing foundational aspects of quantum mechanics and suggest that Hermiticity may emerge dynamically along the semiclassical branch describing our universe.

\end{abstract}

\maketitle

Hermiticity is often regarded as a foundational principle of quantum mechanics, and for good reason, as it guarantees real spectra for physical observables, ensures the conservation of probability through unitary time evolution in closed systems, and underpins a consistent measurement theory via the Born rule within the standard Hilbert-space inner product \cite{qm1bohm2013quantum,qm2zettili2009quantum,qm3sakurai2020modern,qm4griffiths2018introduction,qm5shankar2012principles,qm6scherrer2024quantum}. Nevertheless, non-Hermitian quantum mechanics has developed into an increasingly mature and experimentally relevant framework over the last few decades motivated by the ubiquity of effective descriptions in which degrees of freedom are traced out by design or by inaccessibility \cite{nh1moiseyev2011non,nh2ashida2020non,nh3hatano1996localization}. In such settings, complex potentials and gain loss structures arise naturally and can be implemented controllably in engineered platforms where balanced amplification and absorption or post selected dynamics permit direct laboratory access to non-unitary evolution. This has led to a broad program in which non-Hermitian models are treated both as effective theories of open dynamics and as candidates for generalized fundamental dynamics equipped with modified inner products and biorthogonal structures \cite{nh4jones2014relativistic,nh5gopalakrishnan2021entanglement,nh6hatano1997vortex,nh7bender2007making,nh8longhi2010optical}. As a result, a growing class of quantum experiments has been proposed and realized to constrain effective non-Hermitian generators through precision interferometry, spectral response measurements, non-unitary decay envelopes, and consistency tests of probability conservation in appropriately defined norms \cite{nh9jones2010non,nh10krejvcivrik2015pseudospectra,nh11cui2012geometric,nh12bergholtz2021exceptional}.

A minimal departure from standard quantum mechanics begins by allowing the Hamiltonian to fail to be self adjoint under the canonical inner product while retaining the Schr\"odinger form of time evolution,
\begin{equation}
i\hbar \frac{\partial}{\partial t}\ket{\psi(t)}=\hat H \ket{\psi(t)}
\label{eq:NH_Schro}
\end{equation}
The non-Hermitian Hamiltonian may be decomposed into Hermitian and anti Hermitian parts,
\begin{equation}
\hat H=\hat H_{\rm H}+i\hat \Gamma
\label{eq:H_decomp}
\end{equation}
with $\hat H_{\rm H}=\hat H_{\rm H}^{\dagger}$ and $\hat \Gamma=\hat \Gamma^{\dagger}$. The norm with respect to the standard inner product then evolves according to,
\begin{equation}
\begin{split}
    \frac{d}{dt}\braket{\psi(t)|\psi(t)}&=\frac{i}{\hbar}\bra{\psi(t)}\Big(\hat H^{\dagger}-\hat H\Big)\ket{\psi(t)}\\&=-\frac{2}{\hbar}\bra{\psi(t)}\hat \Gamma\ket{\psi(t)}
\end{split}
\label{eq:norm_evol}
\end{equation}
so that $\hat \Gamma$ acts as a gain or loss generator in the reduced description. For an eigenstate $\hat H\ket{n_{\rm R}}=E_n\ket{n_{\rm R}}$ with $E_n=E_n^{\rm R}+iE_n^{\rm I}$ one obtains amplitudes $e^{-iE_n t/\hbar}=e^{-iE_n^{\rm R}t/\hbar}e^{E_n^{\rm I}t/\hbar}$ so that the imaginary component controls decay or amplification. In such theories the spectral problem is generically biorthogonal and one introduces left and right eigenvectors defined by,
\begin{equation}
\hat H\ket{n_{\rm R}}=E_n\ket{n_{\rm R}},
\enspace
\hat H^{\dagger}\ket{n_{\rm L}}=E_n^{\ast}\ket{n_{\rm L}},
\enspace
\braket{n_{\rm L}|m_{\rm R}}=\delta_{nm}
\label{eq:biorth}
\end{equation}
Expectation values consistent with this structure are naturally written in terms of left and right states as $\langle \hat O\rangle=\bra{\psi_{\rm L}}\hat O\ket{\psi_{\rm R}}$. In a distinguished subclass one may restore a unitary interpretation by modifying the inner product through a positive metric operator $\eta$ such that $\hat H^{\dagger}=\eta \hat H \eta^{-1}$. In that case, the time evolution generated by $\hat H$ can be made norm preserving in the $\eta$ inner product even though it appears non-Hermitian in the naive one. This motivates treating non-Hermitian quantum mechanics as either a fundamental generalization of quantum theory or as an effective description capturing coarse graining conditioning and environment induced irreversibility, and in either interpretation it provides a sharp parametrization of departures from standard unitarity that can be constrained experimentally through the size and functional form of $\hat\Gamma$ \cite{nh13gardas2016non,nh14matsoukas2023non,nh15bender2007faster,nh16cao2023statistical,nh17giri2009non,nh18ju2019non,nh19ju2024emergent}.

Moving beyond just simple quantum mechanics for particles, quantum cosmology presents a whole different way of extending the canonical quantization paradigm to the gravitational field itself, by promoting the spatial geometry and matter configurations to quantum variables described by a wave functional $\Psi$ over the space of three geometries and matter field profiles \cite{qc1bojowald2015quantum,qc2bojowald2008loop,qc4wiltshire1996introduction}. The central notion is that the universe admits a quantum state even in the absence of an external observer, and that semiclassical cosmology should emerge in appropriate limits through WKB or Born-Oppenheimer expansions in which the gravitational degrees of freedom play the role of heavy variables and matter or perturbative modes play the role of light variables. In that regime the universal wave function can lead to an emergent classical spacetime together with effective Schr\"odinger evolution for subsystems and can reproduce the standard Friedmann dynamics as the leading Hamilton-Jacobi limit, while providing a framework for addressing initial conditions, tunneling proposals, and potential singularity resolution \cite{qc5ashtekar2011loop,qc6hawking1987quantum,qc7gell1996quantum,qc8bojowald2011quantum,qc9vilenkin1995predictions,qc10calcagni2017classical}. The conceptual economy is that cosmology and gravity inherit the probabilistic structure of quantum mechanics while the observed classical universe arises as an approximation controlled by the smallness of $\hbar$ compared to the relevant action scales and by decoherence between macroscopically distinct branches. 

One typically develops the framework of quantum cosmology by first rewriting general relativity in Hamiltonian form using the ADM decomposition, in which spacetime is foliated by spatial hypersurfaces and the dynamics is encoded in constraint equations rather than a standard Hamiltonian evolution \cite{adm1Arnowitt:1959ah,adm2DeWitt:1967yk,adm3Arnowitt:1962hi}. The lapse and shift appear as Lagrange multipliers enforcing the Hamiltonian and momentum constraints, and the canonical variables are the induced spatial metric and its conjugate momentum.  Quantization is then achieved by promoting the canonical momenta to functional derivatives acting on a wave functional of the three metric and matter fields, while imposing the constraints as operator conditions annihilating the state. This leads to the Wheeler-DeWitt equation, which has the formal structure of a Klein Gordon type equation on superspace and is analyzed by identifying classically allowed regions where WKB solutions oscillate and classically forbidden regions where solutions are exponential. Boundary condition proposals such as the tunneling prescription motivate particular combinations of these solutions, and in minisuperspace truncations one can study tunneling from small scale factor and the emergence of semiclassical time in the Born Oppenheimer limit as well \cite{qc11ashtekar2009loop,qc12vilenkin1988quantum,qc13vilenkin1994approaches,qc14banerjee2012introduction,qc15halliwell1989decoherence,qc16halliwell1991introductory}. This program provides a controlled arena to connect quantum initial conditions to late-time classical cosmology, while sharpening the conceptual issues associated with time probability and measurement in a closed universe \cite{qc17chataignier2023observations,qc18vilenkin1985classical,qc19moniz2010quantum,qc20pawlowski2012positive,qc21linde1991inflation,qc22anninos2024remarks,qc23pinto2013quantum}. 

Now, let's see what would happen if we take into account non-Hermitian effects into this setup. We begin from the Einstein-Hilbert action with a cosmological constant and minimally coupled matter,
\begin{equation}
S=\frac{M_{\rm Pl}^2}{2}\int d^4x \sqrt{-g}\Big(R-2\Lambda\Big)+S_{\rm m}[g_{\mu\nu},\varphi]
\label{eq:EH_action}
\end{equation}
In ADM variables, the metric is written as $ds^2=-N^2dt^2+h_{ij}(dx^i+N^i dt)(dx^j+N^j dt)$ with $h_{ij}$ the induced three metric and $N$ and $N^i$ the lapse and shift. The action becomes a constrained Hamiltonian system of the form,
\begin{equation}
S=\int dt \int d^3x \Big(\pi^{ij}\dot h_{ij}+\pi_{\varphi}\dot\varphi-N\mathcal H-N^i\mathcal H_i\Big)
\label{eq:ADM_action}
\end{equation}
where $\pi^{ij}$ and $\pi_{\varphi}$ are the canonical conjugate momenta to $h_{ij}$ and $\varphi$. If we then go ahead by varying with respect to $N$ and $N^i$, then it gives us the Hamiltonian and momentum constraints, 
$\mathcal H\approx 0$ and 
$\mathcal H_i\approx 0$.
The explicit structure of the Hamiltonian constraint may be written as,
\begin{equation}
\begin{split}
    \mathcal H&=\frac{1}{M_{\rm Pl}^2\sqrt{h}}\Big(\pi_{ij}\pi^{ij}-\frac{1}{2}\pi^2\Big)-\frac{M_{\rm Pl}^2}{2}\sqrt{h}\Big({}^{(3)}R-2\Lambda\Big)\\&+\mathcal H_{\rm m}[h_{ij},\varphi,\pi_{\varphi}]
\end{split}
\label{eq:H_constraint_explicit}
\end{equation}
with $\pi=h_{ij}\pi^{ij}$ and ${}^{(3)}R$ the Ricci scalar of the spatial metric. The momentum constraint $\mathcal H_i$ enforces spatial diffeomorphism invariance and will be imposed alongside $\mathcal H$ in the quantum theory. The canonical quantization step promotes the phase space variables to operators acting on a wave functional $\Psi[h_{ij},\varphi]$ and in standard quantum cosmology one takes,
\begin{equation}
\pi^{ij}(x)\rightarrow -i\hbar \frac{\delta}{\delta h_{ij}(x)}
\qquad
\pi_{\varphi}(x)\rightarrow -i\hbar \frac{\delta}{\delta \varphi(x)}
\label{eq:std_ops}
\end{equation}
so that the constraints become operator equations annihilating the state. In particular, we see that the Hamiltonian constraint gives the Wheeler-DeWitt equation,
\begin{equation}
\hat{\mathcal H}\Psi[h_{ij},\varphi]=0
\label{eq:WdW}
\end{equation}
which encodes the absence of an external time parameter and implements time reparameterization invariance at the quantum level.

To incorporate non-Hermitian effects, one may generalize the operator realization of the constraint so that the quantum generator on superspace need not be self adjoint under the naive inner product on wave functionals. We therefore define a non-Hermitian Hamiltonian constraint operator by analogy with Eq. \eqref{eq:H_decomp},
\begin{equation}
\hat{\mathcal H}_{\rm NH}=\hat{\mathcal H}_{\rm H}+i\hat \Gamma_{\mathcal H}
\label{eq:NH_constraint_decomp}
\end{equation}
where $\hat{\mathcal H}_{\rm H}=\hat{\mathcal H}_{\rm H}^{\dagger}$ and $\hat \Gamma_{\mathcal H}=\hat \Gamma_{\mathcal H}^{\dagger}$ are operators on superspace  The non-Hermitian Wheeler-DeWitt equation is then postulated as the quantum implementation of the classical constraint
\begin{equation}
\hat{\mathcal H}_{\rm NH}\Psi[h_{ij},\varphi]=0
\label{eq:NH_WdW}
\end{equation}
The interpretation is that $\hat{\mathcal H}_{\rm H}$ encodes the usual kinetic and potential structure inherited from Eq. \eqref{eq:H_constraint_explicit} including factor ordering and measure choices, while $\hat \Gamma_{\mathcal H}$ parameterizes an intrinsic gain loss functional on superspace that may arise effectively through a fundamental departure from Hermiticity. The analogy with Eq.\eqref{eq:NH_Schro} is structural rather than temporal because Eq.\eqref{eq:NH_WdW} is a constraint equation rather than an evolution equation. Nevertheless, the presence of an anti-Hermitian component modifies the structure of conserved currents on superspace and biases the weighting of semiclassical histories relative to the corresponding Hermitian theory.

To make contact with cosmology one often truncates to minisuperspace variables $q^A$ such as the scale factor $a$ and one or more homogeneous matter fields. In that case, the Hermitian Wheeler-DeWitt operator can be written as a Laplace-Beltrami operator on minisuperspace, plus a real potential $U_R(q)$ and matter operators while the non-Hermitian extension naturally corresponds to a complexified potential or more general anti-Hermitian differential term. A representative form is,
\begin{equation}
\left[-\hbar^2\nabla^2_{(G)}+U_R(q)+iU_I(q)\right]\Psi(q)=0
\label{eq:minisup_NH_WdW}
\end{equation}
where $\nabla^2_{(G)}$ is defined by the minisuperspace metric $G_{AB}$ inherited from the kinetic term in Eq.\eqref{eq:H_constraint_explicit} and $U_I$ is the minisuperspace representation of $\hat \Gamma_{\mathcal H}$. In a semiclassical WKB treatment one writes $\Psi\sim \exp(iS/\hbar)$ in classically allowed regions and finds at leading order a Hamilton-Jacobi equation for $S$ modified by the imaginary contribution, which can act as a source or sink for the WKB current. In Born-Oppenheimer type expansions one may further factorize the wave function into heavy gravitational and light matter sectors and the anti-Hermitian part induces non-unitary terms in the emergent Schr\"odinger equation for the light sector in direct analogy with Eq.\eqref{eq:norm_evol}. In this way Eqs. \eqref{eq:NH_constraint_decomp} through \eqref{eq:minisup_NH_WdW} define a minimal and systematic framework for ``Non-Hermitian Quantum Cosmology'', in which departures from standard quantum cosmology are encoded by $\hat \Gamma_{\mathcal H}$ and can be confronted with the observed near consistency between geometric probes and growth probes by translating $\hat \Gamma_{\mathcal H}$ into effective non-unitary rates for cosmological perturbations in the semiclassical limit.

Now, let us see how this non-Hermitian quantum cosmology could be confronted with observations of late-time universe, where the background geometry is well described by an FLRW spacetime and the relevant observables split naturally into geometric probes of the expansion history $H(z)$ and dynamical probes of the growth of structure. In the semiclassical Born-Oppenheimer limit of Eq. \eqref{eq:NH_WdW} or Eq. \eqref{eq:minisup_NH_WdW}, one writes $\Psi\simeq e^{iS_0(q)/\hbar}\psi$ with $S_0$ determining an emergent time variable along a classical background trajectory and the light sector $\psi$ obeys an effective non-unitary Schr\"odinger equation whose structure follows Eq. \eqref{eq:NH_Schro}, whose anti-Hermitian contribution is inherited from $\hat \Gamma_{\mathcal H}$. Denoting the emergent time anti-Hermitian generator by $\hat K(t)$, one has the norm evolution,
\begin{equation}
\frac{d}{dt}\langle \psi|\psi\rangle=-\frac{2}{\hbar}\langle \psi|\hat K(t)|\psi\rangle
\label{eq:late_norm_re}
\end{equation}
which shows that $\hat K$ acts as a sink or source in the reduced description. For late-time large scale structure the observationally relevant imprint of such non-unitary evolution is most directly captured at the level of linear perturbations through an effective modification of the growth of the matter density contrast $\delta_m$. In GR, the subhorizon linear equation in cosmic time is,
\begin{equation}
\ddot \delta_m+2H\dot \delta_m-4\pi G\rho_m \delta_m=0
\label{eq:delta_GR}
\end{equation}
and defining the growth factor by $\delta_m({\bf x},t)=D(t)\delta_m({\bf x},t_{\rm i})$ gives us the standard equation for $D$. In the presence of an anti-Hermitian contribution in the emergent dynamics, the most conservative late-time parametrization that preserves the same background $H(t)$ while allowing non-unitary damping or gain of the perturbations is to add a real friction term $\gamma(t)\dot \delta_m$, giving
\begin{equation}
\ddot \delta_m+\left(2H+\gamma(t)\right)\dot \delta_m-4\pi G\rho_m \delta_m=0
\label{eq:delta_gamma}
\end{equation}
and hence we have
\begin{equation}
\ddot D+\left(2H+\gamma(t)\right)\dot D-4\pi G\rho_m D=0
\label{eq:D_gamma}
\end{equation}
The mapping between $\gamma(t)$ and $\hat K(t)$ is model dependent, but Eq. \eqref{eq:late_norm_re} guarantees that any residual anti-Hermitian component generically induces an exponential envelope in reduced amplitudes and Eq. \eqref{eq:D_gamma} is the minimal way to encode that envelope in the classicalized growth sector.

To extract a bound directly from observations, write the solution as a small deformation of the standard Hermitian cosmology solution $D(t)=D_0(t)e^{\epsilon(t)}$, where $D_0$ solves Eq. \eqref{eq:D_gamma} with $\gamma=0$ and $\epsilon$ is assumed perturbatively small over the relevant redshift interval. One then has
\begin{equation}
\dot D=D_0 e^{\epsilon}\left(\dot \epsilon+\frac{\dot D_0}{D_0}\right)
\label{eq:Ddot_eps}
\end{equation}
\begin{equation}
\ddot D=D_0 e^{\epsilon}\left(\ddot \epsilon+2\dot \epsilon\frac{\dot D_0}{D_0}+\dot \epsilon^2+\frac{\ddot D_0}{D_0}\right)
\label{eq:Dddot_eps}
\end{equation}
Substituting Eqs. \eqref{eq:Ddot_eps} and \eqref{eq:Dddot_eps} into Eq. \eqref{eq:D_gamma}, dividing by $D_0 e^{\epsilon}$, and using that $D_0$ satisfies the baseline equation $\ddot D_0+2H\dot D_0-4\pi G\rho_m D_0=0$, one finds an exact evolution equation for $\epsilon$,
\begin{equation}
\ddot \epsilon+\dot \epsilon^2+2\dot \epsilon\frac{\dot D_0}{D_0}+\gamma(t)\left(\dot \epsilon+\frac{\dot D_0}{D_0}\right)=0
\label{eq:eps_exact}
\end{equation}
In the regime of interest we keep only the leading terms linear in $\epsilon$ and its derivatives, giving us
\begin{equation}
\ddot \epsilon+\left(2\frac{\dot D_0}{D_0}+\gamma(t)\right)\dot \epsilon+\gamma(t)\frac{\dot D_0}{D_0}=0
\label{eq:eps_lin}
\end{equation}
It is also very convenient to define $u(t)=\dot \epsilon(t)$ as then Eq. \eqref{eq:eps_lin} then becomes a first order inhomogeneous equation,
\begin{equation}
\dot u+\left(2\frac{\dot D_0}{D_0}+\gamma(t)\right)u=-\gamma(t)\frac{\dot D_0}{D_0}
\label{eq:u_eq}
\end{equation}
The integrating factor is $\mu(t)=\exp\left[\int^t dt'\left(2\dot D_0/D_0+\gamma\right)\right]=D_0^2(t)\exp\left[\int^t \gamma(t')dt'\right]$, so the solution is
\begin{equation}
u(t)\mu(t)-u(t_{\rm i})\mu(t_{\rm i})=-\int_{t_{\rm i}}^{t} dt'\,\gamma(t')\frac{\dot D_0(t')}{D_0(t')}\,\mu(t')
\label{eq:u_sol_formal}
\end{equation}
Assuming the deformation vanishes at early times in the sense $u(t_{\rm i})\simeq 0$ and using $\dot D_0/D_0=fH$ with $f=d\ln D_0/d\ln a$, one obtains
\begin{equation}
u(t)=-\mu^{-1}(t)\int_{t_{\rm i}}^{t} dt'\,\gamma(t') f(t')H(t')\,\mu(t')
\label{eq:u_sol}
\end{equation}
Since $\mu(t')$ grows rapidly with $D_0^2(t')$ while $\mu^{-1}(t)$ suppresses earlier contributions, the dominant late-time effect is controlled by the integrated history of $\gamma$ weighted over the epoch where growth is active. For slowly varying $\gamma$ and $f$ on Hubble timescales, Eq. \eqref{eq:u_sol} implies the parametric estimate $u\sim -\gamma/2$ and hence,
\begin{equation}
\epsilon(t_0)=\int_{t_{\rm i}}^{t_0} u(t)\,dt\simeq -\frac{1}{2}\int_{t_{\rm i}}^{t_0}\gamma(t)\,dt
\label{eq:eps_est}
\end{equation}
which gives the fractional growth modification,
\begin{equation}
\frac{\Delta D}{D}\simeq \epsilon(t_0)\simeq -\frac{1}{2}\int_{t_{\rm i}}^{t_0}\gamma(t)\,dt
\label{eq:DD_final}
\end{equation}
In linear theory, the clustering amplitude on $8\,h^{-1}{\rm Mpc}$ scales is,
\begin{equation}
\sigma_8^2(t)=\int_0^{\infty}\frac{dk}{2\pi^2}k^2 P_m(k,t)W^2(kR_8)
\label{eq:s8_def}
\end{equation}
with $P_m(k,t)=D^2(t)P_m(k,t_{\rm i})$ for scale independent growth, so that $\sigma_8(t)\propto D(t)$ and therefore,
\begin{equation}
\frac{\Delta \sigma_8}{\sigma_8}\simeq \frac{\Delta D}{D}\simeq -\frac{1}{2}\int_{t_{\rm i}}^{t_0}\gamma(t)\,dt
\label{eq:s8_final}
\end{equation}
Requiring that the non-unitary damping or gain does not spoil the observed near-consistency between growth observables and geometric probes motivates the bound $|\Delta \sigma_8/\sigma_8|\lesssim \varepsilon$ over the redshift range where structure is measured, which yields the integral constraint
\begin{equation}
\left|\int_{t_{\rm i}}^{t_0}\gamma(t)\,dt\right|\lesssim 2\varepsilon
\label{eq:gamma_int_bound_re}
\end{equation}
If $\gamma$ is approximately constant over a characteristic duration $\Delta t$ of order the Hubble time, $\Delta t\sim H_0^{-1}$ then this becomes,
\begin{equation}
\frac{|\gamma_0|}{H_0}\lesssim 2\varepsilon
\label{eq:gamma_simple_bound_re}
\end{equation}
which shows explicitly that any anti-Hermitian imprint feeding the observed late-time growth sector must correspond to a rate far below the natural gravitational timescale set by $H_0$.

Complementary restrictions can also arise from the geometry sector as a  generic non-Hermitian imprint that survives coarse-graining can also act as an effective source or sink for the energy density of a dark component, which may be parametrized by a modified continuity equation of the form,
\begin{equation}
\dot\rho_{\rm DE}+3H\left(1+w\right)\rho_{\rm DE}=Q(t)
\label{eq:contQ_re}
\end{equation}
If $Q$ is nonzero over a Hubble time, it induces a fractional shift in $\rho_{\rm DE}$ of order $\Delta\rho_{\rm DE}/\rho_{\rm DE}\sim \int (Q/\rho_{\rm DE})dt$ and therefore a shift in $H(z)$ through the Friedmann equation. Taking the minimal scaling $Q=\xi H\rho_{\rm DE}$ implies $\rho_{\rm DE}\propto a^{-3(1+w)+\xi}$ and thus an order $\xi$ distortion of $H(z)$ relative to the Hermitian baseline, so the empirical success of precision distance measures requires $|\xi|\ll 1$ over the probed redshift range.

These results provide an important translation between non-Hermitian quantum cosmology and late-time observables. Starting from the non-Hermitian constraint structure in Eq. \eqref{eq:NH_WdW}, the semiclassical reduction gives us an anti-Hermitian generator $\hat K$ whose presence implies non-unitary evolution as in Eq. \eqref{eq:late_norm_re}. When projected onto the classicalized perturbation sector, the leading imprint may be encoded by an effective friction or gain term $\gamma(t)$ in the growth equation \eqref{eq:D_gamma}, which integrates over cosmic time to produce a fractional shift in the growth factor and hence in $\sigma_8$ given by Eqs. \eqref{eq:DD_final} and \eqref{eq:s8_final}. The observed near consistency between geometry and growth constrains the integrated non-unitary rate to satisfy Eq. \eqref{eq:gamma_int_bound_re}, implying the parametric suppression $|\gamma|/H\ll 1$ in the late-time universe. At the same time, precision measurements of $H(z)$ similarly demand that any effective source term in the background sector be small. In physical terms, if the fundamental quantum dynamics of the universe admitted a generic non-Hermitian component that remained unsuppressed at late times, it would induce order unity amplification or attenuation of the growth of structure when integrated over a Hubble time. The absence of such a signal implies that either the non-Hermitian functional $\hat \Gamma_{\mathcal H}$ is confined to high curvature epochs, is safe by an appropriate physical inner product or is dynamically suppressed along the semiclassical branch describing our universe. This elevates cosmology into a novel infrared test of Hermiticity, complementary to laboratory probes, because it constrains not only instantaneous deviations from unitarity but also their accumulated effects over gigayear timescales through the combined requirement that distances and growth be simultaneously described by a single consistent cosmological history.

An additional and independent late-time handle on non-Hermitian quantum cosmology is provided by the observed near flatness of the universe. In an FLRW background, the Friedmann equation reads
\begin{equation}
H^2=\frac{8\pi G}{3}\rho_{\rm tot}-\frac{k}{a^2}
\label{eq:friedmann_curv}
\end{equation}
and the dimensionless curvature parameter is defined as
\begin{equation}
\Omega_K\equiv -\frac{k}{a^2H^2}.
\label{eq:omegaK_def}
\end{equation}
Observationally one finds $|\Omega_K|\ll 1$ at late times, which goes towards implying that the curvature term remains subdominant compared to the energy density driving the expansion. In standard Hermitian quantum cosmology, this is consistent with inflationary attractor behavior and with the classical stability of $k=0$ solutions. In a non-Hermitian setting, however, the Wheeler-DeWitt current on minisuperspace is no longer conserved and the weighting of semiclassical histories can acquire source or sink terms that generically bias trajectories away from finely balanced configurations such as exact spatial flatness.

To see this explicitly, consider a minisuperspace truncation in which the Wheeler-DeWitt equation takes the non-Hermitian form,
\begin{equation}
\left[-\hbar^2\frac{\partial^2}{\partial a^2}+U_R(a,k)+iU_I(a,k)\right]\Psi(a)=0
\label{eq:WdW_curv_NH}
\end{equation}
where $U_R$ contains the usual curvature contribution $-k a^2$ and $U_I$ encodes the non-Hermitian functional inherited from $\hat\Gamma_{\mathcal H}$ and the associated WKB current $J_a$ obeys a modified continuity equation,
\begin{equation}
\frac{dJ_a}{da}=\frac{2}{\hbar}U_I(a,k)|\Psi(a)|^2
\label{eq:WdW_current}
\end{equation}
so that unless $U_I$ vanishes or is tuned to be curvature independent, different curvature sectors experience differential amplification or suppression in superspace. Exact flatness $k=0$ then corresponds to a measure zero trajectory requiring both $U_R(a,0)=0$ and $U_I(a,0)=0$ to avoid exponential bias, while any residual curvature dependence in $U_I$ generically drives probability flow toward $k\neq 0$ branches. 

Projecting this effect into the semiclassical late-time dynamics, a curvature dependent non-Hermitian contribution may be parametrized phenomenologically as an effective evolution equation for $\Omega_K$,
\begin{equation}
\frac{d\Omega_K}{d\ln a}=-2\Omega_K+\xi_K
\label{eq:omegaK_evol}
\end{equation}
where $\xi_K$ represents the integrated effect of the non-Hermitian source or sink on curvature weighting. In the Hermitian limit $\xi_K=0$ and $\Omega_K\propto a^{-2}$ decays in an expanding universe and if $\xi_K\neq 0$, the solution is,
\begin{equation}
\Omega_K(a)=\Omega_K(a_{\rm i})a^{-2}+\frac{\xi_K}{2}\left(1-a^{-2}\right)
\label{eq:omegaK_sol}
\end{equation}
so that even a small constant $\xi_K$ drives $\Omega_K$ toward an asymptotic value of order $\xi_K$ and requiring $|\Omega_K|\ll 1$ today implies,
\begin{equation}
|\xi_K|\ll 1
\label{eq:xiK_bound}
\end{equation}
with the bound strengthened by the fact that $\xi_K$ represents an accumulated non-unitary bias integrated over many e-folds of expansion and in this sense, the observed near flatness of the universe provides a constraint on non-Hermiticity that is complementary to those derived from growth and expansion history. It means that any anti-Hermitian contribution that distinguishes between curvature sectors must be extremely suppressed along the semiclassical branch describing our universe. Together with the bounds from $H(z)$ and $\sigma_8$, near flatness reinforces the conclusion that late-time cosmology tolerates only parametrically small non-Hermitian effects, elevating spatial flatness into yet another infrared consistency condition on the fundamental Hermiticity of quantum mechanics.

We can also discuss non-Hermiticity from the perspective of early-time cosmology, where the imprint of quantum dynamics is most direct because primordial fluctuations originate as vacuum fluctuations of quantum fields on an evolving background. As in the late-time case, we begin from the non-Hermitian Wheeler–DeWitt equation Eq \eqref{eq:NH_WdW} and its minisuperspace realization Eq \eqref{eq:minisup_NH_WdW}. In the semiclassical Born–Oppenheimer expansion one writes $\Psi(q,\chi)\simeq e^{iS_0(q)/\hbar}\psi(q,\chi)$, where $q$ denotes background gravitational variables and $\chi$ collectively denotes perturbative degrees of freedom. The leading order Hamilton–Jacobi equation for $S_0$ defines an emergent time variable $t$ along a classical background trajectory, while the next order gives an effective Schrödinger equation for $\psi$ of the form Eq \eqref{eq:NH_Schro}, supplemented by an anti-Hermitian contribution inherited from $\hat\Gamma_{\mathcal H}$. Denoting the resulting non-unitary generator in the perturbation sector by $\hat K(t)$, the reduced dynamics satisfies the norm evolution Eq \eqref{eq:late_norm_re} indicating that any residual non-Hermiticity generically induces exponential damping or amplification of reduced amplitudes.

For scalar perturbations during inflation, the quadratic action in the Hermitian theory leads to the Mukhanov–Sasaki equation,
\begin{equation}
v_k''+\left(k^2-\frac{z''}{z}\right)v_k=0
\label{eq:MS_std_final}
\end{equation}
where $v_k$ is the canonical variable, $z=a\dot\phi/H$, and primes denote derivatives with respect to conformal time $\tau$. In the non-Hermitian extension, the reduced quadratic Hamiltonian for each Fourier mode may be written as,
\begin{equation}
H_k=\frac{1}{2}\left(\pi_k^2+\omega_k^2 v_k^2\right)+i\Gamma_k(\tau)\,\frac{1}{2}\left(v_k\pi_k+\pi_k v_k\right)
\label{eq:Hk_NH_final}
\end{equation}
with $\omega_k^2=k^2-z''/z$ and real $\Gamma_k$ and the Heisenberg equations following from Eq \eqref{eq:Hk_NH_final} are,
\begin{equation}
v_k'=\pi_k+\Gamma_k v_k
\label{eq:vkprime_final}
\end{equation}
\begin{equation}
\pi_k'=-\omega_k^2 v_k+\Gamma_k \pi_k
\label{eq:pikprime_final}
\end{equation}
Eliminating $\pi_k$ yields a second-order equation for $v_k$,
\begin{equation}
v_k''-2\Gamma_k v_k'+\left(\omega_k^2-\Gamma_k'-\Gamma_k^2\right)v_k=0
\label{eq:MS_NH_full_final}
\end{equation}
For slowly varying $\Gamma_k$ and $|\Gamma_k|\ll \omega_k$, the dominant modification is the first derivative term. Writing $\Gamma_k=-\Gamma$ with $\Gamma>0$ corresponding to damping, Eq. \eqref{eq:MS_NH_full_final} reduces to,
\begin{equation}
v_k''+2\Gamma v_k'+\left(k^2-\frac{z''}{z}\right)v_k\simeq 0
\label{eq:MS_damped_final}
\end{equation}
The solution can be written as $v_k(\tau)=e^{-\int^\tau \Gamma d\tau'}\tilde v_k(\tau)$, where $\tilde v_k$ satisfies the Hermitian equation Eq \eqref{eq:MS_std_final}. The curvature perturbation $\mathcal R_k=v_k/z$ acquires an exponential envelope and the power spectrum becomes,
\begin{equation}
\mathcal P_{\mathcal R}(k)=\mathcal P_{\mathcal R}^{(0)}(k)\exp\left[-2\int^{\tau_k}\Gamma(\tau)\,d\tau\right]
\label{eq:PR_tau_def}
\end{equation}
Changing variables from conformal time to the number of e-folds $N=\ln a$ using $d\tau=(aH)^{-1}dN$ and defining the dimensionless non-Hermitian rate
\begin{equation}
\alpha(N)\equiv \frac{\Gamma(N)}{H(N)}
\label{eq:alpha_def}
\end{equation}
Eq \eqref{eq:PR_tau_def} may be written as,
\begin{equation}
\mathcal P_{\mathcal R}(k)=\mathcal P_{\mathcal R}^{(0)}(k)\exp\left[-2\int^{N_k}\alpha(N)\,dN\right]
\label{eq:PR_alpha}
\end{equation}
where $N_k$ denotes the e-fold at which mode $k$ exits the horizon. Differentiating Eq \eqref{eq:PR_alpha} with respect to $\ln k\simeq N$ gives the modified scalar spectral index,
\begin{equation}
n_s-1\equiv \frac{d\ln \mathcal P_{\mathcal R}}{d\ln k}=\left(n_s-1\right)_0-2\alpha(N_k)
\label{eq:ns_alpha}
\end{equation}
Taking one further derivative gives the running as,
\begin{equation}
\alpha_s\equiv \frac{dn_s}{d\ln k}=\alpha_s^{(0)}-2\frac{d\alpha}{dN}
\label{eq:running_alpha_final}
\end{equation}
An identical construction applies to tensor perturbations with $\omega_k^2=k^2-a''/a$ and a potentially distinct rate $\alpha_T(N)$. The tensor-to-scalar ratio at the pivot scale then becomes,
\begin{equation}
r\equiv \frac{\mathcal P_T}{\mathcal P_{\mathcal R}}=r_0\exp\left[-2\int^{N_\ast}\left(\alpha_T-\alpha\right)dN\right]
\label{eq:r_alpha}
\end{equation}
These expressions make explicit how non-Hermitian contributions in the Wheeler–DeWitt operator propagate through the semiclassical expansion into effective damping terms for primordial fluctuations. Because the observable CMB window spans several e-folds, any non-Hermitian rate $\alpha$ that is not parametrically small would be exponentially amplified across this interval leading to order unity distortions of $\mathcal P_{\mathcal R}(k)$, a large shift in $n_s$ and a sizable running $\alpha_s$, all of which are observationally excluded. Consistency with Eqs. \eqref{eq:ns_alpha}, \eqref{eq:running_alpha_final} and \eqref{eq:r_alpha} therefore requires $\alpha\ll 1$ and $d\alpha/dN\ll 1$ during horizon exit. When combined with the late-time bounds derived from Eqs \eqref{eq:DD_final} and \eqref{eq:s8_final}, this demonstrates a coherent picture in which the same non-Hermitian functional $\hat\Gamma_{\mathcal H}$ is suppressed along the entire semiclassical history of the universe. Early accelerated expansion and late-time structure formation thus act as complementary filters enforcing effective Hermiticity in the infrared, while leaving open the possibility that non-Hermitian structures play a role only in the deep ultraviolet or in regimes where the classical notion of spacetime itself ceases to apply.

A natural next question is whether the strong suppression of late-time and inflationary non-Hermitian effects inferred above is a special feature of the Einstein–Hilbert dynamics or whether it may be partially relaxed once the gravitational sector is generalized beyond GR. A minimal modification is to include higher curvature operators in the effective action, for example,
\begin{equation}
S=\frac{M_{\rm Pl}^2}{2}\int d^4x\sqrt{-g}\Big(R-2\Lambda+\alpha R^2\Big)+S_{\rm m}[g_{\mu\nu},\varphi]
\label{eq:fR_action}
\end{equation}
which can be viewed as the first correction in a curvature expansion of the low-energy effective field theory of gravity. Introducing an auxiliary field $\chi$ yields an equivalent scalar–tensor representation,
\begin{equation}
S=\int d^4x\sqrt{-g}\left[\frac{M_{\rm Pl}^2}{2}\chi R-\frac{M_{\rm Pl}^2}{2}U(\chi)\right]+S_{\rm m}
\label{eq:aux_action}
\end{equation}
where $U(\chi)$ is determined by the choice of $f(R)$ and where $\chi$ becomes dynamical in the Einstein frame as an additional scalar degree of freedom. In a minisuperspace truncation, this enlargement of the gravitational sector implies that the Wheeler–DeWitt wave function depends on an extended set of configuration variables, $\Psi\rightarrow \Psi(q^A)$ with $q^A=(a,\chi,\ldots)$ and the Hermitian part of the Hamiltonian constraint takes the form,
\begin{equation}
\hat{\mathcal H}_{\rm H}\rightarrow -\hbar^2\nabla^2_{(G)}+U_{R^2}(a,\chi)+\hat H_{\rm m}
\label{eq:WdW_modgrav}
\end{equation}
where both the minisuperspace metric $G_{AB}$ and the real potential $U_{R^2}$ differ from the Einstein–Hilbert case through the presence of $\chi$ and its potential. Allowing for non-Hermiticity as in Eq. \eqref{eq:NH_constraint_decomp} then gives us,
\begin{equation}
\hat{\mathcal H}_{\rm NH}=\hat{\mathcal H}_{\rm H}+i\hat \Gamma_{\mathcal H}
\qquad
\hat{\mathcal H}_{\rm NH}\Psi=0
\label{eq:NH_modgrav}
\end{equation}
so that the same anti-Hermitian functional $\hat \Gamma_{\mathcal H}$ now acts on an enlarged superspace. In the semiclassical reduction, the late-time growth equation Eq. \eqref{eq:D_gamma} is generically replaced by a form with an effective gravitational coupling and modified friction sourced by the additional gravitational degree of freedom,
\begin{equation}
\ddot D+\left(2H+\gamma(t)\right)\dot D-4\pi G_{\rm eff}(t,k)\rho_m D=0
\label{eq:MG_growth}
\end{equation}
where $G_{\rm eff}$ encodes the modification of the gravitational clustering strength and the same non-unitary rate $\gamma$ arises from the anti-Hermitian component in the reduced dynamics. This illustrates how going beyond GR can partially accommodate non-Hermitian effects as a positive shift in $G_{\rm eff}$ can counteract non-unitary damping $\gamma>0$ in the growth sector, while modifications of the background dynamics can compensate small source terms such as Eq \eqref{eq:contQ_re} in the geometry sector. Equivalently, we see that the observational constraints derived in the GR limit do not bound $\gamma$ in isolation but rather constrain combinations of $\gamma$, $H(z)$ and $G_{\rm eff}$, so that the allowed region in parameter space can widen once additional gravitational operators are present.

The deeper implication is that the apparent degree of Hermiticity required by cosmological observations can depend on the gravitational effective field theory used to interpret the data. In the infrared, cosmology constrains a closed and remarkably consistent semiclassical history, enforcing that any effective anti-Hermitian rates remain small when projected onto observable sectors, as quantified by Eqs. \eqref{eq:gamma_simple_bound_re}, \eqref{eq:ns_alpha} and \eqref{eq:r_alpha}. In the ultraviolet  however, higher curvature operators such as those in Eq.~\eqref{eq:fR_action} are precisely the operators expected from integrating out heavy degrees of freedom in quantum gravity, and the same coarse-graining that generates them can also generate effective non-unitarity in a reduced description. From this perspective,  it is striking that a modification of a purely gravitational action can render the observational tolerance on a quintessentially quantum property, Hermiticity, somewhat more flexible, suggesting that what is constrained by cosmology is the joint consistency of the gravitational EFT and the effective quantum evolution of accessible degrees of freedom rather than Hermiticity as an isolated axiom. This provides a concrete arena in which ultraviolet physics can influence the apparent strictness of infrared unitarity bounds, and it motivates viewing non-Hermitian quantum cosmology and modified gravity as two correlated shadows of the same underlying quantum-gravitational completion.

In this work it is important to emphasize that the strong suppression of effective non-Hermiticity inferred from cosmological observations should not be interpreted as a fundamental prohibition of non-Hermitian structures in quantum theory, nor as a statement that nonzero spatial curvature or modified gravitational dynamics are incompatible with quantum mechanics. Rather, Hermiticity in the observable sector emerges as a consistency condition enforced along the particular semiclassical branch describing our universe and this condition depends on the gravitational effective field theory used to define the semiclassical expansion and inner product. In Einstein–Hilbert gravity, the structure of superspace, the conservation of the Wheeler–DeWitt current and the absence of curvature-dependent bias collectively enforce extremely tight infrared constraints on any residual anti-Hermitian contributions. Once the gravitational sector is generalized beyond general relativity, however, the resulting enlargement of configuration space, the modification of the minisuperspace measure, and the altered mapping between microscopic evolution and observable amplitudes imply that the same cosmological data constrain only specific combinations of non-Hermitian rates and gravitational parameters, rather than Hermiticity in isolation. From this perspective, effective unitarity and Hermiticity are not standalone axioms but emergent properties whose apparent rigidity in the infrared reflects the remarkable self-consistency of the semiclassical history of our universe, rather than an absolute prohibition on departures from Hermiticity in more general quantum-gravitational frameworks.

In conclusion, we have formulated a minimal framework for non-Hermitian quantum cosmology by extending the Wheeler–DeWitt constraint to include an anti-Hermitian functional, Eqs. \eqref{eq:NH_constraint_decomp} and \eqref{eq:NH_WdW}, and we have shown how the resulting non-unitary effects propagate through the semiclassical Born–Oppenheimer expansion into late-time structure growth and early-universe primordial fluctuations. Late-time probes constrain the integrated non-unitary rate through its impact on $D(t)$ and $\sigma_8$ via Eqs. \eqref{eq:DD_final} and \eqref{eq:s8_final}, near flatness provides an additional infrared consistency condition through Eqs. \eqref{eq:omegaK_def} and \eqref{eq:omegaK_sol}, and early-time data constrain non-Hermiticity during horizon exit through its exponential imprint on $\mathcal P_{\mathcal R}$ and on $n_s$ and $r$ as in Eqs. \eqref{eq:PR_alpha}, \eqref{eq:ns_alpha}, and \eqref{eq:r_alpha}. The agreement between the assessments from early and late-time constraints suggests that the semiclassical branch describing our universe is effectively Hermitian across vast dynamical ranges, while leaving open the possibility that non-Hermitian structures are relevant only in the deep ultraviolet or in regimes where the classical spacetime description fails. It is remarkable that cosmological observations, which are traditionally viewed as probes of gravity and the large-scale universe, can also be interpreted as indirect but powerful constraints on foundational properties of quantum mechanics. This also raises the intriguing possibility that future high precision cosmological data may permit even sharper tests of core quantum-mechanical principles in regimes inaccessible to laboratory experiments.

\noindent \textbf{Acknowledgements:} We gratefully acknowledge support from Vanderbilt University and the U.S. National Science Foundation. The work of AG is supported in part by NSF Award PHY-2411502, and the work of OT is supported in part by the Vanderbilt Discovery Doctoral Fellowship.


\bibliography{apssamp}

\begin{thebibliography}{50}%
\makeatletter
\providecommand \@ifxundefined [1]{%
 \@ifx{#1\undefined}
}%
\providecommand \@ifnum [1]{%
 \ifnum #1\expandafter \@firstoftwo
 \else \expandafter \@secondoftwo
 \fi
}%
\providecommand \@ifx [1]{%
 \ifx #1\expandafter \@firstoftwo
 \else \expandafter \@secondoftwo
 \fi
}%
\providecommand \natexlab [1]{#1}%
\providecommand \enquote  [1]{``#1''}%
\providecommand \bibnamefont  [1]{#1}%
\providecommand \bibfnamefont [1]{#1}%
\providecommand \citenamefont [1]{#1}%
\providecommand \href@noop [0]{\@secondoftwo}%
\providecommand \href [0]{\begingroup \@sanitize@url \@href}%
\providecommand \@href[1]{\@@startlink{#1}\@@href}%
\providecommand \@@href[1]{\endgroup#1\@@endlink}%
\providecommand \@sanitize@url [0]{\catcode `\\12\catcode `\$12\catcode `\&12\catcode `\#12\catcode `\^12\catcode `\_12\catcode `\%12\relax}%
\providecommand \@@startlink[1]{}%
\providecommand \@@endlink[0]{}%
\providecommand \url  [0]{\begingroup\@sanitize@url \@url }%
\providecommand \@url [1]{\endgroup\@href {#1}{\urlprefix }}%
\providecommand \urlprefix  [0]{URL }%
\providecommand \Eprint [0]{\href }%
\providecommand \doibase [0]{https://doi.org/}%
\providecommand \selectlanguage [0]{\@gobble}%
\providecommand \bibinfo  [0]{\@secondoftwo}%
\providecommand \bibfield  [0]{\@secondoftwo}%
\providecommand \translation [1]{[#1]}%
\providecommand \BibitemOpen [0]{}%
\providecommand \bibitemStop [0]{}%
\providecommand \bibitemNoStop [0]{.\EOS\space}%
\providecommand \EOS [0]{\spacefactor3000\relax}%
\providecommand \BibitemShut  [1]{\csname bibitem#1\endcsname}%
\let\auto@bib@innerbib\@empty
\bibitem [{\citenamefont {B{\"o}hm}(2013)}]{qm1bohm2013quantum}%
  \BibitemOpen
  \bibfield  {author} {\bibinfo {author} {\bibfnamefont {A.}~\bibnamefont {B{\"o}hm}},\ }\href@noop {} {\emph {\bibinfo {title} {Quantum mechanics: foundations and applications}}}\ (\bibinfo  {publisher} {Springer Science \& Business Media},\ \bibinfo {year} {2013})\BibitemShut {NoStop}%
\bibitem [{\citenamefont {Zettili}(2009)}]{qm2zettili2009quantum}%
  \BibitemOpen
  \bibfield  {author} {\bibinfo {author} {\bibfnamefont {N.}~\bibnamefont {Zettili}},\ }\bibfield  {title} {\bibinfo {title} {Quantum mechanics: concepts and applications},\ }\href@noop {} {\  (\bibinfo {year} {2009})}\BibitemShut {NoStop}%
\bibitem [{\citenamefont {Sakurai}\ and\ \citenamefont {Napolitano}(2020)}]{qm3sakurai2020modern}%
  \BibitemOpen
  \bibfield  {author} {\bibinfo {author} {\bibfnamefont {J.~J.}\ \bibnamefont {Sakurai}}\ and\ \bibinfo {author} {\bibfnamefont {J.}~\bibnamefont {Napolitano}},\ }\href@noop {} {\emph {\bibinfo {title} {Modern quantum mechanics}}}\ (\bibinfo  {publisher} {Cambridge University Press},\ \bibinfo {year} {2020})\BibitemShut {NoStop}%
\bibitem [{\citenamefont {Griffiths}\ and\ \citenamefont {Schroeter}(2018)}]{qm4griffiths2018introduction}%
  \BibitemOpen
  \bibfield  {author} {\bibinfo {author} {\bibfnamefont {D.~J.}\ \bibnamefont {Griffiths}}\ and\ \bibinfo {author} {\bibfnamefont {D.~F.}\ \bibnamefont {Schroeter}},\ }\href@noop {} {\emph {\bibinfo {title} {Introduction to quantum mechanics}}}\ (\bibinfo  {publisher} {Cambridge university press},\ \bibinfo {year} {2018})\BibitemShut {NoStop}%
\bibitem [{\citenamefont {Shankar}(2012)}]{qm5shankar2012principles}%
  \BibitemOpen
  \bibfield  {author} {\bibinfo {author} {\bibfnamefont {R.}~\bibnamefont {Shankar}},\ }\href@noop {} {\emph {\bibinfo {title} {Principles of quantum mechanics}}}\ (\bibinfo  {publisher} {Springer Science \& Business Media},\ \bibinfo {year} {2012})\BibitemShut {NoStop}%
\bibitem [{\citenamefont {Scherrer}(2024)}]{qm6scherrer2024quantum}%
  \BibitemOpen
  \bibfield  {author} {\bibinfo {author} {\bibfnamefont {R.~J.}\ \bibnamefont {Scherrer}},\ }\href@noop {} {\emph {\bibinfo {title} {Quantum mechanics: an accessible introduction}}}\ (\bibinfo  {publisher} {World Scientific},\ \bibinfo {year} {2024})\BibitemShut {NoStop}%
\bibitem [{\citenamefont {Moiseyev}(2011)}]{nh1moiseyev2011non}%
  \BibitemOpen
  \bibfield  {author} {\bibinfo {author} {\bibfnamefont {N.}~\bibnamefont {Moiseyev}},\ }\href@noop {} {\emph {\bibinfo {title} {Non-Hermitian quantum mechanics}}}\ (\bibinfo  {publisher} {Cambridge University Press},\ \bibinfo {year} {2011})\BibitemShut {NoStop}%
\bibitem [{\citenamefont {Ashida}\ \emph {et~al.}(2020)\citenamefont {Ashida}, \citenamefont {Gong},\ and\ \citenamefont {Ueda}}]{nh2ashida2020non}%
  \BibitemOpen
  \bibfield  {author} {\bibinfo {author} {\bibfnamefont {Y.}~\bibnamefont {Ashida}}, \bibinfo {author} {\bibfnamefont {Z.}~\bibnamefont {Gong}},\ and\ \bibinfo {author} {\bibfnamefont {M.}~\bibnamefont {Ueda}},\ }\bibfield  {title} {\bibinfo {title} {Non-hermitian physics},\ }\href@noop {} {\bibfield  {journal} {\bibinfo  {journal} {Advances in Physics}\ }\textbf {\bibinfo {volume} {69}},\ \bibinfo {pages} {249} (\bibinfo {year} {2020})}\BibitemShut {NoStop}%
\bibitem [{\citenamefont {Hatano}\ and\ \citenamefont {Nelson}(1996)}]{nh3hatano1996localization}%
  \BibitemOpen
  \bibfield  {author} {\bibinfo {author} {\bibfnamefont {N.}~\bibnamefont {Hatano}}\ and\ \bibinfo {author} {\bibfnamefont {D.~R.}\ \bibnamefont {Nelson}},\ }\bibfield  {title} {\bibinfo {title} {Localization transitions in non-hermitian quantum mechanics},\ }\href@noop {} {\bibfield  {journal} {\bibinfo  {journal} {Physical review letters}\ }\textbf {\bibinfo {volume} {77}},\ \bibinfo {pages} {570} (\bibinfo {year} {1996})}\BibitemShut {NoStop}%
\bibitem [{\citenamefont {Jones-Smith}\ and\ \citenamefont {Mathur}(2014)}]{nh4jones2014relativistic}%
  \BibitemOpen
  \bibfield  {author} {\bibinfo {author} {\bibfnamefont {K.}~\bibnamefont {Jones-Smith}}\ and\ \bibinfo {author} {\bibfnamefont {H.}~\bibnamefont {Mathur}},\ }\bibfield  {title} {\bibinfo {title} {Relativistic non-hermitian quantum mechanics},\ }\href@noop {} {\bibfield  {journal} {\bibinfo  {journal} {Physical Review D}\ }\textbf {\bibinfo {volume} {89}},\ \bibinfo {pages} {125014} (\bibinfo {year} {2014})}\BibitemShut {NoStop}%
\bibitem [{\citenamefont {Gopalakrishnan}\ and\ \citenamefont {Gullans}(2021)}]{nh5gopalakrishnan2021entanglement}%
  \BibitemOpen
  \bibfield  {author} {\bibinfo {author} {\bibfnamefont {S.}~\bibnamefont {Gopalakrishnan}}\ and\ \bibinfo {author} {\bibfnamefont {M.~J.}\ \bibnamefont {Gullans}},\ }\bibfield  {title} {\bibinfo {title} {Entanglement and purification transitions in non-hermitian quantum mechanics},\ }\href@noop {} {\bibfield  {journal} {\bibinfo  {journal} {Physical review letters}\ }\textbf {\bibinfo {volume} {126}},\ \bibinfo {pages} {170503} (\bibinfo {year} {2021})}\BibitemShut {NoStop}%
\bibitem [{\citenamefont {Hatano}\ and\ \citenamefont {Nelson}(1997)}]{nh6hatano1997vortex}%
  \BibitemOpen
  \bibfield  {author} {\bibinfo {author} {\bibfnamefont {N.}~\bibnamefont {Hatano}}\ and\ \bibinfo {author} {\bibfnamefont {D.~R.}\ \bibnamefont {Nelson}},\ }\bibfield  {title} {\bibinfo {title} {Vortex pinning and non-hermitian quantum mechanics},\ }\href@noop {} {\bibfield  {journal} {\bibinfo  {journal} {Physical Review B}\ }\textbf {\bibinfo {volume} {56}},\ \bibinfo {pages} {8651} (\bibinfo {year} {1997})}\BibitemShut {NoStop}%
\bibitem [{\citenamefont {Bender}(2007)}]{nh7bender2007making}%
  \BibitemOpen
  \bibfield  {author} {\bibinfo {author} {\bibfnamefont {C.~M.}\ \bibnamefont {Bender}},\ }\bibfield  {title} {\bibinfo {title} {Making sense of non-hermitian hamiltonians},\ }\href@noop {} {\bibfield  {journal} {\bibinfo  {journal} {Reports on Progress in Physics}\ }\textbf {\bibinfo {volume} {70}},\ \bibinfo {pages} {947} (\bibinfo {year} {2007})}\BibitemShut {NoStop}%
\bibitem [{\citenamefont {Longhi}(2010)}]{nh8longhi2010optical}%
  \BibitemOpen
  \bibfield  {author} {\bibinfo {author} {\bibfnamefont {S.}~\bibnamefont {Longhi}},\ }\bibfield  {title} {\bibinfo {title} {Optical realization of relativistic non-hermitian quantum mechanics},\ }\href@noop {} {\bibfield  {journal} {\bibinfo  {journal} {Physical review letters}\ }\textbf {\bibinfo {volume} {105}},\ \bibinfo {pages} {013903} (\bibinfo {year} {2010})}\BibitemShut {NoStop}%
\bibitem [{\citenamefont {Jones-Smith}(2010)}]{nh9jones2010non}%
  \BibitemOpen
  \bibfield  {author} {\bibinfo {author} {\bibfnamefont {K.~A.}\ \bibnamefont {Jones-Smith}},\ }\emph {\bibinfo {title} {Non-Hermitian quantum mechanics}},\ \href@noop {} {Ph.D. thesis},\ \bibinfo  {school} {Case Western Reserve University} (\bibinfo {year} {2010})\BibitemShut {NoStop}%
\bibitem [{\citenamefont {Krej{\v{c}}i{\v{r}}{\'\i}k}\ \emph {et~al.}(2015)\citenamefont {Krej{\v{c}}i{\v{r}}{\'\i}k}, \citenamefont {Siegl}, \citenamefont {Tater},\ and\ \citenamefont {Viola}}]{nh10krejvcivrik2015pseudospectra}%
  \BibitemOpen
  \bibfield  {author} {\bibinfo {author} {\bibfnamefont {D.}~\bibnamefont {Krej{\v{c}}i{\v{r}}{\'\i}k}}, \bibinfo {author} {\bibfnamefont {P.}~\bibnamefont {Siegl}}, \bibinfo {author} {\bibfnamefont {M.}~\bibnamefont {Tater}},\ and\ \bibinfo {author} {\bibfnamefont {J.}~\bibnamefont {Viola}},\ }\bibfield  {title} {\bibinfo {title} {Pseudospectra in non-hermitian quantum mechanics},\ }\href@noop {} {\bibfield  {journal} {\bibinfo  {journal} {Journal of mathematical physics}\ }\textbf {\bibinfo {volume} {56}} (\bibinfo {year} {2015})}\BibitemShut {NoStop}%
\bibitem [{\citenamefont {Cui}\ and\ \citenamefont {Zheng}(2012)}]{nh11cui2012geometric}%
  \BibitemOpen
  \bibfield  {author} {\bibinfo {author} {\bibfnamefont {X.-D.}\ \bibnamefont {Cui}}\ and\ \bibinfo {author} {\bibfnamefont {Y.}~\bibnamefont {Zheng}},\ }\bibfield  {title} {\bibinfo {title} {Geometric phases in non-hermitian quantum mechanics},\ }\href@noop {} {\bibfield  {journal} {\bibinfo  {journal} {Physical Review A—Atomic, Molecular, and Optical Physics}\ }\textbf {\bibinfo {volume} {86}},\ \bibinfo {pages} {064104} (\bibinfo {year} {2012})}\BibitemShut {NoStop}%
\bibitem [{\citenamefont {Bergholtz}\ \emph {et~al.}(2021)\citenamefont {Bergholtz}, \citenamefont {Budich},\ and\ \citenamefont {Kunst}}]{nh12bergholtz2021exceptional}%
  \BibitemOpen
  \bibfield  {author} {\bibinfo {author} {\bibfnamefont {E.~J.}\ \bibnamefont {Bergholtz}}, \bibinfo {author} {\bibfnamefont {J.~C.}\ \bibnamefont {Budich}},\ and\ \bibinfo {author} {\bibfnamefont {F.~K.}\ \bibnamefont {Kunst}},\ }\bibfield  {title} {\bibinfo {title} {Exceptional topology of non-hermitian systems},\ }\href@noop {} {\bibfield  {journal} {\bibinfo  {journal} {Reviews of Modern Physics}\ }\textbf {\bibinfo {volume} {93}},\ \bibinfo {pages} {015005} (\bibinfo {year} {2021})}\BibitemShut {NoStop}%
\bibitem [{\citenamefont {Gardas}\ \emph {et~al.}(2016)\citenamefont {Gardas}, \citenamefont {Deffner},\ and\ \citenamefont {Saxena}}]{nh13gardas2016non}%
  \BibitemOpen
  \bibfield  {author} {\bibinfo {author} {\bibfnamefont {B.}~\bibnamefont {Gardas}}, \bibinfo {author} {\bibfnamefont {S.}~\bibnamefont {Deffner}},\ and\ \bibinfo {author} {\bibfnamefont {A.}~\bibnamefont {Saxena}},\ }\bibfield  {title} {\bibinfo {title} {Non-hermitian quantum thermodynamics},\ }\href@noop {} {\bibfield  {journal} {\bibinfo  {journal} {Scientific reports}\ }\textbf {\bibinfo {volume} {6}},\ \bibinfo {pages} {23408} (\bibinfo {year} {2016})}\BibitemShut {NoStop}%
\bibitem [{\citenamefont {Matsoukas-Roubeas}\ \emph {et~al.}(2023)\citenamefont {Matsoukas-Roubeas}, \citenamefont {Roccati}, \citenamefont {Cornelius}, \citenamefont {Xu}, \citenamefont {Chenu},\ and\ \citenamefont {del Campo}}]{nh14matsoukas2023non}%
  \BibitemOpen
  \bibfield  {author} {\bibinfo {author} {\bibfnamefont {A.~S.}\ \bibnamefont {Matsoukas-Roubeas}}, \bibinfo {author} {\bibfnamefont {F.}~\bibnamefont {Roccati}}, \bibinfo {author} {\bibfnamefont {J.}~\bibnamefont {Cornelius}}, \bibinfo {author} {\bibfnamefont {Z.}~\bibnamefont {Xu}}, \bibinfo {author} {\bibfnamefont {A.}~\bibnamefont {Chenu}},\ and\ \bibinfo {author} {\bibfnamefont {A.}~\bibnamefont {del Campo}},\ }\bibfield  {title} {\bibinfo {title} {Non-hermitian hamiltonian deformations in quantum mechanics},\ }\href@noop {} {\bibfield  {journal} {\bibinfo  {journal} {Journal of High Energy Physics}\ }\textbf {\bibinfo {volume} {2023}},\ \bibinfo {pages} {1} (\bibinfo {year} {2023})}\BibitemShut {NoStop}%
\bibitem [{\citenamefont {Bender}\ \emph {et~al.}(2007)\citenamefont {Bender}, \citenamefont {Brody}, \citenamefont {Jones},\ and\ \citenamefont {Meister}}]{nh15bender2007faster}%
  \BibitemOpen
  \bibfield  {author} {\bibinfo {author} {\bibfnamefont {C.~M.}\ \bibnamefont {Bender}}, \bibinfo {author} {\bibfnamefont {D.~C.}\ \bibnamefont {Brody}}, \bibinfo {author} {\bibfnamefont {H.~F.}\ \bibnamefont {Jones}},\ and\ \bibinfo {author} {\bibfnamefont {B.~K.}\ \bibnamefont {Meister}},\ }\bibfield  {title} {\bibinfo {title} {Faster than hermitian quantum mechanics},\ }\href@noop {} {\bibfield  {journal} {\bibinfo  {journal} {Physical Review Letters}\ }\textbf {\bibinfo {volume} {98}},\ \bibinfo {pages} {040403} (\bibinfo {year} {2007})}\BibitemShut {NoStop}%
\bibitem [{\citenamefont {Cao}\ and\ \citenamefont {Kou}(2023)}]{nh16cao2023statistical}%
  \BibitemOpen
  \bibfield  {author} {\bibinfo {author} {\bibfnamefont {K.}~\bibnamefont {Cao}}\ and\ \bibinfo {author} {\bibfnamefont {S.-P.}\ \bibnamefont {Kou}},\ }\bibfield  {title} {\bibinfo {title} {Statistical mechanics for non-hermitian quantum systems},\ }\href@noop {} {\bibfield  {journal} {\bibinfo  {journal} {Physical Review Research}\ }\textbf {\bibinfo {volume} {5}},\ \bibinfo {pages} {033196} (\bibinfo {year} {2023})}\BibitemShut {NoStop}%
\bibitem [{\citenamefont {Giri}\ and\ \citenamefont {Roy}(2009)}]{nh17giri2009non}%
  \BibitemOpen
  \bibfield  {author} {\bibinfo {author} {\bibfnamefont {P.~R.}\ \bibnamefont {Giri}}\ and\ \bibinfo {author} {\bibfnamefont {P.}~\bibnamefont {Roy}},\ }\bibfield  {title} {\bibinfo {title} {Non-hermitian quantum mechanics in non-commutative space},\ }\href@noop {} {\bibfield  {journal} {\bibinfo  {journal} {The European Physical Journal C}\ }\textbf {\bibinfo {volume} {60}},\ \bibinfo {pages} {157} (\bibinfo {year} {2009})}\BibitemShut {NoStop}%
\bibitem [{\citenamefont {Ju}\ \emph {et~al.}(2019)\citenamefont {Ju}, \citenamefont {Miranowicz}, \citenamefont {Chen},\ and\ \citenamefont {Nori}}]{nh18ju2019non}%
  \BibitemOpen
  \bibfield  {author} {\bibinfo {author} {\bibfnamefont {C.-Y.}\ \bibnamefont {Ju}}, \bibinfo {author} {\bibfnamefont {A.}~\bibnamefont {Miranowicz}}, \bibinfo {author} {\bibfnamefont {G.-Y.}\ \bibnamefont {Chen}},\ and\ \bibinfo {author} {\bibfnamefont {F.}~\bibnamefont {Nori}},\ }\bibfield  {title} {\bibinfo {title} {Non-hermitian hamiltonians and no-go theorems in quantum information},\ }\href@noop {} {\bibfield  {journal} {\bibinfo  {journal} {Physical Review A}\ }\textbf {\bibinfo {volume} {100}},\ \bibinfo {pages} {062118} (\bibinfo {year} {2019})}\BibitemShut {NoStop}%
\bibitem [{\citenamefont {Ju}\ \emph {et~al.}(2024)\citenamefont {Ju}, \citenamefont {Miranowicz}, \citenamefont {Chen}, \citenamefont {Chen},\ and\ \citenamefont {Nori}}]{nh19ju2024emergent}%
  \BibitemOpen
  \bibfield  {author} {\bibinfo {author} {\bibfnamefont {C.-Y.}\ \bibnamefont {Ju}}, \bibinfo {author} {\bibfnamefont {A.}~\bibnamefont {Miranowicz}}, \bibinfo {author} {\bibfnamefont {Y.-N.}\ \bibnamefont {Chen}}, \bibinfo {author} {\bibfnamefont {G.-Y.}\ \bibnamefont {Chen}},\ and\ \bibinfo {author} {\bibfnamefont {F.}~\bibnamefont {Nori}},\ }\bibfield  {title} {\bibinfo {title} {Emergent parallel transport and curvature in hermitian and non-hermitian quantum mechanics},\ }\href@noop {} {\bibfield  {journal} {\bibinfo  {journal} {Quantum}\ }\textbf {\bibinfo {volume} {8}},\ \bibinfo {pages} {1277} (\bibinfo {year} {2024})}\BibitemShut {NoStop}%
\bibitem [{\citenamefont {Bojowald}(2015)}]{qc1bojowald2015quantum}%
  \BibitemOpen
  \bibfield  {author} {\bibinfo {author} {\bibfnamefont {M.}~\bibnamefont {Bojowald}},\ }\bibfield  {title} {\bibinfo {title} {Quantum cosmology: a review},\ }\href@noop {} {\bibfield  {journal} {\bibinfo  {journal} {Reports on Progress in Physics}\ }\textbf {\bibinfo {volume} {78}},\ \bibinfo {pages} {023901} (\bibinfo {year} {2015})}\BibitemShut {NoStop}%
\bibitem [{\citenamefont {Bojowald}(2008)}]{qc2bojowald2008loop}%
  \BibitemOpen
  \bibfield  {author} {\bibinfo {author} {\bibfnamefont {M.}~\bibnamefont {Bojowald}},\ }\bibfield  {title} {\bibinfo {title} {Loop quantum cosmology},\ }\href@noop {} {\bibfield  {journal} {\bibinfo  {journal} {Living Reviews in Relativity}\ }\textbf {\bibinfo {volume} {11}},\ \bibinfo {pages} {4} (\bibinfo {year} {2008})}\BibitemShut {NoStop}%
\bibitem [{\citenamefont {Wiltshire}\ \emph {et~al.}(1996)\citenamefont {Wiltshire} \emph {et~al.}}]{qc4wiltshire1996introduction}%
  \BibitemOpen
  \bibfield  {author} {\bibinfo {author} {\bibfnamefont {D.~L.}\ \bibnamefont {Wiltshire}} \emph {et~al.},\ }\bibfield  {title} {\bibinfo {title} {An introduction to quantum cosmology},\ }\href@noop {} {\bibfield  {journal} {\bibinfo  {journal} {Cosmology: the Physics of the Universe}\ ,\ \bibinfo {pages} {473}} (\bibinfo {year} {1996})}\BibitemShut {NoStop}%
\bibitem [{\citenamefont {Ashtekar}\ and\ \citenamefont {Singh}(2011)}]{qc5ashtekar2011loop}%
  \BibitemOpen
  \bibfield  {author} {\bibinfo {author} {\bibfnamefont {A.}~\bibnamefont {Ashtekar}}\ and\ \bibinfo {author} {\bibfnamefont {P.}~\bibnamefont {Singh}},\ }\bibfield  {title} {\bibinfo {title} {Loop quantum cosmology: a status report},\ }\href@noop {} {\bibfield  {journal} {\bibinfo  {journal} {Classical and Quantum Gravity}\ }\textbf {\bibinfo {volume} {28}},\ \bibinfo {pages} {213001} (\bibinfo {year} {2011})}\BibitemShut {NoStop}%
\bibitem [{\citenamefont {Hawking}(1987)}]{qc6hawking1987quantum}%
  \BibitemOpen
  \bibfield  {author} {\bibinfo {author} {\bibfnamefont {S.~W.}\ \bibnamefont {Hawking}},\ }\bibfield  {title} {\bibinfo {title} {Quantum cosmology.},\ }\href@noop {} {\bibfield  {journal} {\bibinfo  {journal} {Three Hundred Years of Gravitation}\ ,\ \bibinfo {pages} {631}} (\bibinfo {year} {1987})}\BibitemShut {NoStop}%
\bibitem [{\citenamefont {Gell-Mann}\ and\ \citenamefont {Hartle}(1996)}]{qc7gell1996quantum}%
  \BibitemOpen
  \bibfield  {author} {\bibinfo {author} {\bibfnamefont {M.}~\bibnamefont {Gell-Mann}}\ and\ \bibinfo {author} {\bibfnamefont {J.~B.}\ \bibnamefont {Hartle}},\ }\bibfield  {title} {\bibinfo {title} {Quantum mechanics in the light of quantum cosmology},\ }in\ \href@noop {} {\emph {\bibinfo {booktitle} {Foundations of Quantum Mechanics in the Light of New Technology: Selected Papers from the Proceedings of the First through Fourth International Symposia on Foundations of Quantum Mechanics}}}\ (\bibinfo {organization} {World Scientific},\ \bibinfo {year} {1996})\ pp.\ \bibinfo {pages} {347--369}\BibitemShut {NoStop}%
\bibitem [{\citenamefont {Bojowald}(2011)}]{qc8bojowald2011quantum}%
  \BibitemOpen
  \bibfield  {author} {\bibinfo {author} {\bibfnamefont {M.}~\bibnamefont {Bojowald}},\ }\href@noop {} {\emph {\bibinfo {title} {Quantum cosmology}}}\ (\bibinfo  {publisher} {Springer},\ \bibinfo {year} {2011})\BibitemShut {NoStop}%
\bibitem [{\citenamefont {Vilenkin}(1995)}]{qc9vilenkin1995predictions}%
  \BibitemOpen
  \bibfield  {author} {\bibinfo {author} {\bibfnamefont {A.}~\bibnamefont {Vilenkin}},\ }\bibfield  {title} {\bibinfo {title} {Predictions from quantum cosmology},\ }\href@noop {} {\bibfield  {journal} {\bibinfo  {journal} {Physical Review Letters}\ }\textbf {\bibinfo {volume} {74}},\ \bibinfo {pages} {846} (\bibinfo {year} {1995})}\BibitemShut {NoStop}%
\bibitem [{\citenamefont {Calcagni}(2017)}]{qc10calcagni2017classical}%
  \BibitemOpen
  \bibfield  {author} {\bibinfo {author} {\bibfnamefont {G.}~\bibnamefont {Calcagni}},\ }\href@noop {} {\emph {\bibinfo {title} {Classical and quantum cosmology}}}\ (\bibinfo  {publisher} {Springer},\ \bibinfo {year} {2017})\BibitemShut {NoStop}%
\bibitem [{\citenamefont {Arnowitt}\ \emph {et~al.}(1959)\citenamefont {Arnowitt}, \citenamefont {Deser},\ and\ \citenamefont {Misner}}]{adm1Arnowitt:1959ah}%
  \BibitemOpen
  \bibfield  {author} {\bibinfo {author} {\bibfnamefont {R.~L.}\ \bibnamefont {Arnowitt}}, \bibinfo {author} {\bibfnamefont {S.}~\bibnamefont {Deser}},\ and\ \bibinfo {author} {\bibfnamefont {C.~W.}\ \bibnamefont {Misner}},\ }\bibfield  {title} {\bibinfo {title} {{Dynamical Structure and Definition of Energy in General Relativity}},\ }\href {https://doi.org/10.1103/PhysRev.116.1322} {\bibfield  {journal} {\bibinfo  {journal} {Phys. Rev.}\ }\textbf {\bibinfo {volume} {116}},\ \bibinfo {pages} {1322} (\bibinfo {year} {1959})}\BibitemShut {NoStop}%
\bibitem [{\citenamefont {DeWitt}(1967)}]{adm2DeWitt:1967yk}%
  \BibitemOpen
  \bibfield  {author} {\bibinfo {author} {\bibfnamefont {B.~S.}\ \bibnamefont {DeWitt}},\ }\bibfield  {title} {\bibinfo {title} {{Quantum Theory of Gravity. 1. The Canonical Theory}},\ }\href {https://doi.org/10.1103/PhysRev.160.1113} {\bibfield  {journal} {\bibinfo  {journal} {Phys. Rev.}\ }\textbf {\bibinfo {volume} {160}},\ \bibinfo {pages} {1113} (\bibinfo {year} {1967})}\BibitemShut {NoStop}%
\bibitem [{\citenamefont {Arnowitt}\ \emph {et~al.}(2008)\citenamefont {Arnowitt}, \citenamefont {Deser},\ and\ \citenamefont {Misner}}]{adm3Arnowitt:1962hi}%
  \BibitemOpen
  \bibfield  {author} {\bibinfo {author} {\bibfnamefont {R.~L.}\ \bibnamefont {Arnowitt}}, \bibinfo {author} {\bibfnamefont {S.}~\bibnamefont {Deser}},\ and\ \bibinfo {author} {\bibfnamefont {C.~W.}\ \bibnamefont {Misner}},\ }\bibfield  {title} {\bibinfo {title} {{The Dynamics of general relativity}},\ }\href {https://doi.org/10.1007/s10714-008-0661-1} {\bibfield  {journal} {\bibinfo  {journal} {Gen. Rel. Grav.}\ }\textbf {\bibinfo {volume} {40}},\ \bibinfo {pages} {1997} (\bibinfo {year} {2008})},\ \Eprint {https://arxiv.org/abs/gr-qc/0405109} {arXiv:gr-qc/0405109} \BibitemShut {NoStop}%
\bibitem [{\citenamefont {Ashtekar}(2009)}]{qc11ashtekar2009loop}%
  \BibitemOpen
  \bibfield  {author} {\bibinfo {author} {\bibfnamefont {A.}~\bibnamefont {Ashtekar}},\ }\bibfield  {title} {\bibinfo {title} {Loop quantum cosmology: an overview},\ }\href@noop {} {\bibfield  {journal} {\bibinfo  {journal} {General Relativity and Gravitation}\ }\textbf {\bibinfo {volume} {41}},\ \bibinfo {pages} {707} (\bibinfo {year} {2009})}\BibitemShut {NoStop}%
\bibitem [{\citenamefont {Vilenkin}(1988)}]{qc12vilenkin1988quantum}%
  \BibitemOpen
  \bibfield  {author} {\bibinfo {author} {\bibfnamefont {A.}~\bibnamefont {Vilenkin}},\ }\bibfield  {title} {\bibinfo {title} {Quantum cosmology and the initial state of the universe},\ }\href@noop {} {\bibfield  {journal} {\bibinfo  {journal} {Physical Review D}\ }\textbf {\bibinfo {volume} {37}},\ \bibinfo {pages} {888} (\bibinfo {year} {1988})}\BibitemShut {NoStop}%
\bibitem [{\citenamefont {Vilenkin}(1994)}]{qc13vilenkin1994approaches}%
  \BibitemOpen
  \bibfield  {author} {\bibinfo {author} {\bibfnamefont {A.}~\bibnamefont {Vilenkin}},\ }\bibfield  {title} {\bibinfo {title} {Approaches to quantum cosmology},\ }\href@noop {} {\bibfield  {journal} {\bibinfo  {journal} {Physical Review D}\ }\textbf {\bibinfo {volume} {50}},\ \bibinfo {pages} {2581} (\bibinfo {year} {1994})}\BibitemShut {NoStop}%
\bibitem [{\citenamefont {Banerjee}\ \emph {et~al.}(2012)\citenamefont {Banerjee}, \citenamefont {Calcagni}, \citenamefont {Martin-Benito} \emph {et~al.}}]{qc14banerjee2012introduction}%
  \BibitemOpen
  \bibfield  {author} {\bibinfo {author} {\bibfnamefont {K.}~\bibnamefont {Banerjee}}, \bibinfo {author} {\bibfnamefont {G.}~\bibnamefont {Calcagni}}, \bibinfo {author} {\bibfnamefont {M.}~\bibnamefont {Martin-Benito}}, \emph {et~al.},\ }\bibfield  {title} {\bibinfo {title} {Introduction to loop quantum cosmology},\ }\href@noop {} {\bibfield  {journal} {\bibinfo  {journal} {SIGMA. Symmetry, Integrability and Geometry: Methods and Applications}\ }\textbf {\bibinfo {volume} {8}},\ \bibinfo {pages} {016} (\bibinfo {year} {2012})}\BibitemShut {NoStop}%
\bibitem [{\citenamefont {Halliwell}(1989)}]{qc15halliwell1989decoherence}%
  \BibitemOpen
  \bibfield  {author} {\bibinfo {author} {\bibfnamefont {J.~J.}\ \bibnamefont {Halliwell}},\ }\bibfield  {title} {\bibinfo {title} {Decoherence in quantum cosmology},\ }\href@noop {} {\bibfield  {journal} {\bibinfo  {journal} {Physical Review D}\ }\textbf {\bibinfo {volume} {39}},\ \bibinfo {pages} {2912} (\bibinfo {year} {1989})}\BibitemShut {NoStop}%
\bibitem [{\citenamefont {Halliwell}(1991)}]{qc16halliwell1991introductory}%
  \BibitemOpen
  \bibfield  {author} {\bibinfo {author} {\bibfnamefont {J.~J.}\ \bibnamefont {Halliwell}},\ }\bibfield  {title} {\bibinfo {title} {Introductory lectures on quantum cosmology},\ }in\ \href@noop {} {\emph {\bibinfo {booktitle} {Quantum cosmology and baby universes}}}\ (\bibinfo  {publisher} {World Scientific},\ \bibinfo {year} {1991})\ pp.\ \bibinfo {pages} {159--243}\BibitemShut {NoStop}%
\bibitem [{\citenamefont {Chataignier}\ \emph {et~al.}(2023)\citenamefont {Chataignier}, \citenamefont {Kiefer},\ and\ \citenamefont {Moniz}}]{qc17chataignier2023observations}%
  \BibitemOpen
  \bibfield  {author} {\bibinfo {author} {\bibfnamefont {L.}~\bibnamefont {Chataignier}}, \bibinfo {author} {\bibfnamefont {C.}~\bibnamefont {Kiefer}},\ and\ \bibinfo {author} {\bibfnamefont {P.}~\bibnamefont {Moniz}},\ }\bibfield  {title} {\bibinfo {title} {Observations in quantum cosmology},\ }\href@noop {} {\bibfield  {journal} {\bibinfo  {journal} {Classical and Quantum Gravity}\ }\textbf {\bibinfo {volume} {40}},\ \bibinfo {pages} {223001} (\bibinfo {year} {2023})}\BibitemShut {NoStop}%
\bibitem [{\citenamefont {Vilenkin}(1985)}]{qc18vilenkin1985classical}%
  \BibitemOpen
  \bibfield  {author} {\bibinfo {author} {\bibfnamefont {A.}~\bibnamefont {Vilenkin}},\ }\bibfield  {title} {\bibinfo {title} {Classical and quantum cosmology of the starobinsky inflationary model},\ }\href@noop {} {\bibfield  {journal} {\bibinfo  {journal} {Physical Review D}\ }\textbf {\bibinfo {volume} {32}},\ \bibinfo {pages} {2511} (\bibinfo {year} {1985})}\BibitemShut {NoStop}%
\bibitem [{\citenamefont {Moniz}(2010)}]{qc19moniz2010quantum}%
  \BibitemOpen
  \bibfield  {author} {\bibinfo {author} {\bibfnamefont {P.~V.}\ \bibnamefont {Moniz}},\ }\href@noop {} {\emph {\bibinfo {title} {Quantum Cosmology-The Supersymmetric Perspective-Vol. 1: Fundamentals}}},\ Vol.\ \bibinfo {volume} {803}\ (\bibinfo  {publisher} {Springer},\ \bibinfo {year} {2010})\BibitemShut {NoStop}%
\bibitem [{\citenamefont {Paw{\l}owski}\ and\ \citenamefont {Ashtekar}(2012)}]{qc20pawlowski2012positive}%
  \BibitemOpen
  \bibfield  {author} {\bibinfo {author} {\bibfnamefont {T.}~\bibnamefont {Paw{\l}owski}}\ and\ \bibinfo {author} {\bibfnamefont {A.}~\bibnamefont {Ashtekar}},\ }\bibfield  {title} {\bibinfo {title} {Positive cosmological constant in loop quantum cosmology},\ }\href@noop {} {\bibfield  {journal} {\bibinfo  {journal} {Physical Review D—Particles, Fields, Gravitation, and Cosmology}\ }\textbf {\bibinfo {volume} {85}},\ \bibinfo {pages} {064001} (\bibinfo {year} {2012})}\BibitemShut {NoStop}%
\bibitem [{\citenamefont {Linde}(1991)}]{qc21linde1991inflation}%
  \BibitemOpen
  \bibfield  {author} {\bibinfo {author} {\bibfnamefont {A.}~\bibnamefont {Linde}},\ }\bibfield  {title} {\bibinfo {title} {Inflation and quantum cosmology},\ }\href@noop {} {\bibfield  {journal} {\bibinfo  {journal} {Physica Scripta}\ }\textbf {\bibinfo {volume} {1991}},\ \bibinfo {pages} {30} (\bibinfo {year} {1991})}\BibitemShut {NoStop}%
\bibitem [{\citenamefont {Anninos}\ \emph {et~al.}(2024)\citenamefont {Anninos}, \citenamefont {Baracco},\ and\ \citenamefont {M{\"u}hlmann}}]{qc22anninos2024remarks}%
  \BibitemOpen
  \bibfield  {author} {\bibinfo {author} {\bibfnamefont {D.}~\bibnamefont {Anninos}}, \bibinfo {author} {\bibfnamefont {C.}~\bibnamefont {Baracco}},\ and\ \bibinfo {author} {\bibfnamefont {B.}~\bibnamefont {M{\"u}hlmann}},\ }\bibfield  {title} {\bibinfo {title} {Remarks on 2d quantum cosmology},\ }\href@noop {} {\bibfield  {journal} {\bibinfo  {journal} {Journal of Cosmology and Astroparticle Physics}\ }\textbf {\bibinfo {volume} {2024}}\bibinfo  {number} { (10)},\ \bibinfo {pages} {031}}\BibitemShut {NoStop}%
\bibitem [{\citenamefont {Pinto-Neto}\ and\ \citenamefont {Fabris}(2013)}]{qc23pinto2013quantum}%
  \BibitemOpen
\bibfield  {number} {  }\bibfield  {author} {\bibinfo {author} {\bibfnamefont {N.}~\bibnamefont {Pinto-Neto}}\ and\ \bibinfo {author} {\bibfnamefont {J.}~\bibnamefont {Fabris}},\ }\bibfield  {title} {\bibinfo {title} {Quantum cosmology from the de broglie--bohm perspective},\ }\href@noop {} {\bibfield  {journal} {\bibinfo  {journal} {Classical and Quantum Gravity}\ }\textbf {\bibinfo {volume} {30}},\ \bibinfo {pages} {143001} (\bibinfo {year} {2013})}\BibitemShut {NoStop}%
\end{thebibliography}%

\end{document}